\documentclass{pasa}%

\title[Multi-Frequency Study of \ngc]{A Multi-Frequency Study of the Milky Way-like Spiral Galaxy \ngc}
\author[M. Yew et al.]{Miranda Yew$^1$, Miroslav D. Filipovi\'c$^1$, Quentin Roper$^1$, Jordan D. Collier$^{1,2}$, Evan J. Crawford$^1$, Thomas H. Jarrett$^{3}$,  Nicholas F. H. Tothill$^1$, Andrew N. O'Brien$^{1,2}$, Marko Z. Pavlovi{\'c}$^{4}$, Thomas G. Pannuti$^{5}$, Timothy J. Galvin$^{1,2,6}$, Anna D. Kapi\'nska$^{7,8}$, Michelle E. Cluver$^{9}$, Julie K. Banfield$^{1,10}$, Eric M. Schlegel$^{11}$, Nigel Maxted$^{1,12}$ \and Kevin R. Grieve$^1$ \\
\\
\affil{$^1$Western Sydney University, Locked Bag 1797, Penrith South DC, NSW, Australia}%
\affil{$^2$CSIRO Astronomy and Space Science (CASS), Marsfield, NSW 2122, Australia}
\affil{$^3$Astrophysics, Cosmology and Gravity Centre (ACGC), Astronomy Department, University of Cape Town, Private Bag X3, Rondebosch 7701, South Africa}
\affil{$^4$Department of Astronomy, Faculty of Mathematics, University of Belgrade, Studentski~trg 16, 11000 Belgrade, Serbia}
\affil{$^5$Space Science Center, Department of Earth and Space Sciences, Morehead State University, 235 Martindale Drive, Morehead, KY 40351, USA}
\affil{$^6$International Centre for Radio Astronomy Research (ICRAR), Curtin University, Bentley, WA 6102, Australia}
\affil{$^7$International Centre for Radio Astronomy Research (ICRAR), University of Western Australia, 35 Stirling Hwy, WA 6009, Australia}%
\affil{$^8$ARC Centre of Excellence for All-Sky Astrophysics (CAASTRO)}
\affil{$^9$Department of Physics and Astronomy, University of the Western Cape, 7535 Bellville, South Africa}
\affil{$^{10}$Research School of Astronomy and Astrophysics, Australian National University, Canberra, ACT 2611, Australia}
\affil{$^{11}$The University of Texas, One UTSA Circle, San Antonio, 78249, Texas}
\affil{$^{12}$School of Physics, The University of New South Wales, Sydney, 2052, Australia}
}%
\jid{PASA}
\doi{10.1017/pas.\the\year.2017}
\jyear{\the\year}

\usepackage[authoryear]{natbib}
\bibpunct{(}{)}{;}{a}{}{,}
\usepackage{aas_macros}
\usepackage{longtable}
\usepackage{rotating}
\usepackage{hyperref}
\hypersetup{colorlinks,citecolor=blue,linkcolor=blue,urlcolor=blue}

\hypersetup{draft}

\newcommand{\einstein}{{\it Einstein}}

\newcommand{\chandra}{{\it Chandra}}

\def\HI{\hbox{H\,{\sc i}}}
\def\HII{\hbox{H\,{\sc ii}}}

\newcommand{\Halpha}{H${\alpha}$}

\newcommand{\ngc}{\mbox {NGC\,6744}}
\def\p0{\phantom{0}}

\newcommand{\D}{$^\circ$}

\def\arcmin{\hbox{$^\prime$}}
\def\arcsec{\hbox{$^{\prime\prime}$}}

\begin{document}%
\begin{abstract}

We present a multi-frequency study of the intermediate spiral SAB(r)bc type galaxy \ngc, using available data from the \chandra\ X-Ray telescope, radio continuum data from the Australia Telescope Compact Array and Murchison Widefield Array, and Wide-field Infrared Survey Explorer infrared observations. We identify 117 X-ray sources and 280 radio sources. Of these, we find nine sources in common between the X-ray and radio catalogues, one of which is a faint central black hole with a bolometric radio luminosity similar to the Milky Way's central black hole. We classify 5 objects as supernova remnant candidates, 2 objects as likely supernova remnants, 17 as \HII\ regions, 1 source as an AGN; the remaining 255 radio sources are categorised as background objects and one X-ray source is classified as a foreground star. We find the star-formation rate (SFR) of \ngc\ to be in the range 2.8 -- 4.7~$\rm{M_{\odot}~yr^{-1}}$ signifying the galaxy is still actively forming stars. The specific SFR of \ngc\ is greater than that of late-type spirals such as the Milky Way, but considerably less that that of a typical starburst galaxy. 
\end{abstract}
\begin{keywords}
galaxy -- \ngc\ -- radio continuum -- X-ray -- infrared -- AGN -- supernovae -- SN2005at
\end{keywords}
\maketitle%

\section{INTRODUCTION}
 \label{sec:intro}

The galaxy \ngc\ \citep[discovered in 1823;][]{1828RSPT..118..113D} is thought to be one of the most Milky Way-like spiral galaxies in our immediate vicinity, with flocculent arms and an elongated core. \ngc\ is classified as an intermediate spiral SAB(r)bc type galaxy and is located in the Pavo-Indus cloud \citep{1975gaun.book..557D}. It sits in a star-rich field, hosts a central ring structure \citep{1963ApJ...138..934D} and is inclined at 50\D $\pm$4\D\ \citep{1999PASA...16...84R} to our line of sight. The only prominent companion of \ngc\ is an irregular dwarf galaxy NGC\,6744A \citep{1999PASA...16...84R} similar to the Milky Way's companions, the Magellanic Clouds.

\citet{1538-3881-146-4-86} estimated a distance to \ngc\ to be $9.2\pm0.4$~Mpc, using the `tip of the red giant branch' (TRGB) method \citep[see][]{1993ApJ...417..553L}. To more easily compare our results with those of previous studies, we adopt the distance of $9.5\pm0.6$~Mpc to \ngc\ as derived by \citet{2014A&A...572A..75K}, so the apparent optical size of 20\arcmin$\times$12.9\arcmin\ corresponds to 55$\times$35~kpc.

Previous studies of \ngc\ include an investigation of the \HI\ disk \citep{1995ApJ...444..610R,1999PASA...16...84R}, \HII\ regions in the optical range \citep[$\lambda=3700\,\textrm{\AA}-6800\,\textrm{\AA}$;][]{1982ApJ...252..594T,1995ApJ...444..610R} and the galaxy's optical shape and size \citep{1963ApJ...138..934D}. \Halpha\ based star formation rate (SFR) indicators suggest a rate of \mbox{3.3--6.8~${\rm{M_{\odot}~{yr^{-1}}}}$} \citep{2012MSAIS..19..158B,1994ApJ...430..142R}. An X-ray study by \citet{1992ApJS...80..531F} analysed an \einstein\ Imaging Proportional Counter image of \ngc\ (see their Figure 7) and identified weak emission. 

\ngc\ is an ideal target for studying the structure, composition and evolutionary tracks of objects such as supernova remnants (SNRs) and \HII\ regions, and comparing them to their counterparts in the Milky Way. The \ngc\ objects, though at various evolutionary phases, can be approximated to be at the same distance to us. This is not the case in the Milky Way where such distances often have large associated fractional errors. \ngc\ is therefore a useful analogue to our own Galaxy when it comes to studying objects such as SNRs and \HII\ regions.

In 2005, the type~Ic supernova SN2005at was discovered in \ngc\ at optical wavebands by \citet{2005CBET..119....1M}, classified by \citet{2005CBET..122....1S}, and studied extensively by \citet{2014A&A...572A..75K}. Co-discovery optical images\footnote{Taken by L.~A.~G. Monard on 5$^{th}$ March 2005.} give a magnitude of 14.3, while \citet{2005CBET..119....1M} measured a magnitude of 16 at the discovery epoch (15$^{th}$ March 2005). Light curve fitting \citep{2014A&A...572A..75K} suggests that the magnitude would have peaked at $\rm{m_R}\sim$14. The  discovery of SN2005at prompted a wide range of multi-frequency observations towards \ngc, which allow us to study the galaxy in detail across the electromagnetic spectrum. 

This paper is a continuation of our studies on nearby galaxies including the LMC \citep[]{2007MNRAS.382..543H}, SMC \citep[]{2005MNRAS.364..217F,2011SerAJ.182...43W,2011SerAJ.183..103W,2011SerAJ.183...95C,2012A&A...545A.128H,2012SerAJ.185...53W,2013A&A...558A...3S}, NGC\,300 \citep[]{2004A&A...425..443P,2011Ap&SS.332..221M,2012Ap&SS.340..133G,2012SerAJ.184...19M}, NGC\,55 \citep[]{Brien2013}, M\,31 \citep[]{2012SerAJ.184...41G,2014SerAJ.189...15G}, NGC\,7793 \citep[]{2011AJ....142...20P,2012MNRAS.427..956D,2014Ap&SS.353..603G} and NGC\,45 \citep[]{1538-3881-150-3-91}. We use archival X-ray data from the \chandra\ X-ray Observatory \citep{2000SPIE.4012....2W} and infrared images obtained with the Wide-field Infrared Survey Explorer \citep[\textit{WISE};][]{wright2010wide}. We examine all available archival radio-continuum observations carried out with the Australia Telescope Compact Array \citep[ATCA;][]{1992JEEEA..12..103F} and the Murchison Widefield Array \citep[MWA;][]{2013PASA...30....7T}. We combine archived ATCA data to create new images with high angular resolution and excellent sensitivity. In Section~\ref{sec:observational data} we describe the observational data and reduction techniques. In Section~\ref{sec:results and discussion} we present our results and discussion and in Section~\ref{sec:conclusion} we summarise our findings.

\section{OBSERVATIONAL DATA}
\label{sec:observational data}

\subsection{X-Ray Data}
\begin{figure*}
 \begin{center}
  \resizebox{1\columnwidth}{!}{\includegraphics[trim = 0 0 0 0]{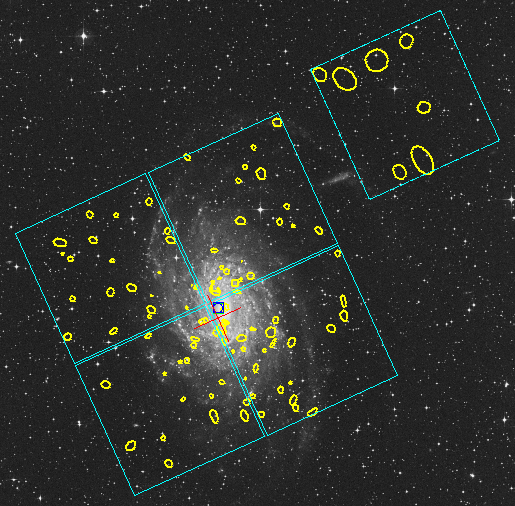}}
  \resizebox{1\columnwidth}{!}{\includegraphics[trim = 0 0 0 0]{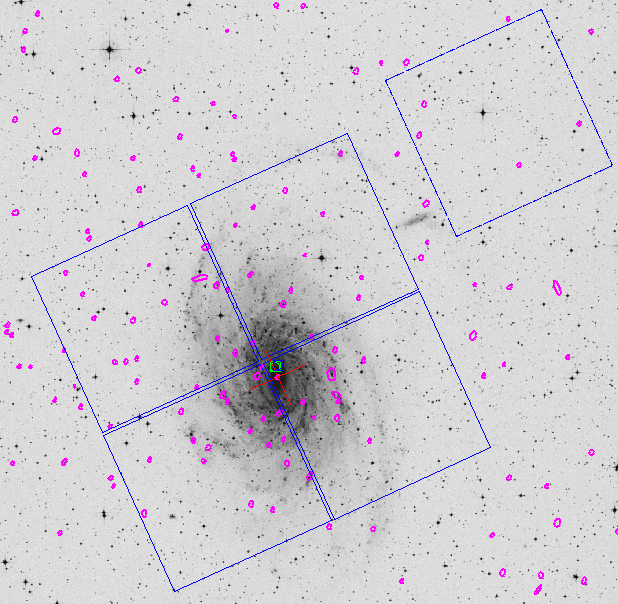}}
  \caption{Five ACIS-I CCD chip footprints overlaid on the optical DSS2 image of \ngc.
The 6$^{th}$ CCD was omitted because of instrument noise. The left image shows X-ray broad-band detected sources in yellow and the right image shows a distribution of detected radio sources reported in this paper. The red cross represents the centre of \ngc, the aimpoint is the green square and the nominal aimpoint is the blue square in the centre of the image. North is up and east is to the left.}
  \label{acisfov}
 \end{center}
\end{figure*}

The X-ray data were obtained by the \chandra\ X-ray Observatory Advanced CCD Imaging Spectrometer \citep[ACIS; PI:][observation ID~15384]{2003acis} on 2014~May~5 with an effective exposure time 52.79~ks. The \chandra\ data cover only a fraction of \ngc's optical extent (see Figure~\ref{acisfov}); the 20\arcmin\ major axis of \ngc\ exceeds the 16\arcmin\ ACIS-I field of view. 

X-ray data reduction used the Chandra Interactive Analysis of Observations 
\citep[CIAO;][]{2006SPIE.6270E..1VF} version 4.8 package with CALDB version 4.7.2. Exposure-corrected images and exposure maps were created from events files using the task {\tt chandra\_repro}. Images in five energy bands were created using the {\tt fluximage} task. There was no need for background flare filtering as the images had little background flaring. The five bands considered in this paper are the following: (1) broad band with energies 0.5--7.0~keV and effective energy 2.3~keV; (2) ultrasoft band with energies 0.2--0.5~keV and effective energy 0.4~keV; (3) soft band with energies 0.5--1.2~keV and effective energy 0.92~keV; (4) medium band with energies 1.2--2.0~keV and effective energy 1.56~keV; and (5) the hard band with energies 2.0--7.0~keV and effective energy 3.8~keV. As the characterization of the ACIS response is not well defined below $\sim 0.5$~keV \citep{1998SPIE.3444..267P}, we have excluded the ultrasoft band from our analysis.


Using the task {\tt mkpsfmap}, the exposure-corrected images were used to produce a point spread function (PSF) map which provides information on the PSF at each pixel in the image. The PSF map along with the exposure-corrected images were then fed through the {\tt wavdetect} tool in CIAO to search for significant correlations with Mexican-hat wavelets of different scales. We found 117 sources using wavelet scales of 1.0, 2.0, 4.0, 8.0; in both the xscale and yscale and fluxscales of 1, 2, 3, 4 with a limiting statistical significance of 3$\sigma$. The great advantage of this technique was that it could separate closely-spaced point sources allowing the identification of many different sources in the field of view. Once the images were obtained, a catalogue was compiled and hardness ratios ($H\!R$) were calculated for each point source. This ratio can give an estimate of a source's X-ray spectral energy distribution to see if it is dominated by thermal or non-thermal process. It is calculated as a normalised difference of the exposure-corrected integrated flux between two energy bands where $E$ is the X-ray intensity:
\begin{equation}
 \label{eq11}
 H\!R_1=\frac{E_{\mathrm{Med}}-E_{\mathrm{Soft}}}{E_{\mathrm{Med}}+E_{\mathrm{Soft}}},
\end{equation} 

\begin{equation}
 \label{eq12}
 H\!R_2=\frac{E_{\mathrm{Hard}}-E_{\mathrm{Med}}}{E_{\mathrm{Hard}}+E_{\mathrm{Med}}}.
\end{equation} 
We calculated two hardness ratios ($H\!R_1$ and $H\!R_2$), between medium and soft bands and between hard and medium bands, respectively. Broad-band fluxes were estimated using a power law model with a photon index $\Gamma=1.7$ and column density $N_{H}$ of 0.061$\times 10^{22}\  \mathrm{atoms~cm}^{-2}$ \citep[via][]{1992ApJS...80..531F} to convert count rates to fluxes using \emph{Sherpa} \citep{2001SPIE.4477...76F}. For the remaining three bands; soft, medium and hard, the flux was estimated from flux-corrected images. Although the $H\!R$s are reliable, the absolute fluxes are not.

\subsection{Radio Data}
The $\lambda$=20~cm, 13~cm, 6~cm and 3~cm ($\nu$=1.4~GHz, 2.1~GHz, 5~GHz and 9~GHz) radio data were obtained from the Australia Telescope Online Archive (ATOA)\footnote{\url{http://atoa.atnf.csiro.au/}}. We used all available ATOA radio-continuum observations of \ngc\ (for details see Table~\ref{images}).

\begin{table*}

 \hspace*{-2cm}
  \caption{ATCA data used in this project, identified by original project ID. Each row lists a separate image/data (or combinations of data from different project IDs), its frequency, bandwidth (BW), array, observing date, synthesised beam and root mean square (r.m.s) values ($1\sigma$). All images were created with a robust (natural) weighting of 2.}
 \vspace{1mm}
  \small
 \begin{center}
\begin{tabular}{lcclccc}
 \hline\hline
  \multicolumn{1}{p{1.8cm}}{ATCA Proj. ID} &
  \multicolumn{1}{p{1cm}}{\centering Freq. \\ (MHz)} &
  \multicolumn{1}{p{1cm}}{\centering BW \\ (MHz)} &
  \multicolumn{1}{p{1cm}}{Array} &
  \multicolumn{1}{p{2.8cm}}{\centering Date(s) \\ (DD/MM/YY) } &
  \multicolumn{1}{p{2.0cm}}{\centering FWHM \\ (\arcsec$\times$\arcsec)} &
  \multicolumn{1}{p{2.0cm}}{\centering r.m.s. \\ (mJy~beam$^{-1}$)} \\
 \hline
C2697      & 2100 & 2048 & 6D       & 15/06/12          &  7.51$\times$4.41   & 0.010 \\
CX082      & 4800 & 128  & 1.5A     & 13/04/05          & 10.43$\times$5.13   & 0.045 \\
           & 8640 & 128  & 6A       & 30/03/05          &  2.68$\times$1.12   & 0.187 \\
C514       & 1380 & 128  & 1.5A     & 10/11/96          & 34.13$\times$31.30  & 0.379 \\
C287       & 4800 & 128  & 375      & 17/08/95          & 20.40$\times$15.80  & 0.831 \\
           & 8640 & 128  & 375      & 17/08/95          & 11.23$\times$8.84   & 0.853 \\
           & 4800 & 128  & 750A     & 28/02/95          & 37.09$\times$34.94  & 0.680 \\
           & 8640 & 128  & 750A     & 28/02/95          & 20.62$\times$19.43  & 0.550 \\
C389       & 1380 & 128  & 1.5D     & 23/09/94          & 37.20$\times$25.13  & 0.082 \\
           & 2380 & 128  & 1.5A     & 23/09/94          & 21.15$\times$14.30  & 0.047 \\
 \hline
CX082+C287 & 4800 & 128  & 1.5A+375 & 13/04/05+17/08/95 & 11.26$\times$5.17   & 0.210 \\
           & 8640 & 128  & 6A+375   & 30/03/05+17/08/95 &  6.23$\times$2.86   & 0.544 \\
C514+C389  & 1380 & 128  & 1.5A     & 13/04/96+23/09/94 & 34.08$\times$31.30  & 0.129 \\
  \hline
\end{tabular}
 \label{images}
 \end{center}
\end{table*}

Two of the seven selected ATCA projects, C184 and C892, were found to have poor $(u,v)$ coverage, resulting in poor images, even after combining compatible projects and implementing joint deconvolution; these data were discarded. The five remaining projects consist of one (C2697) Compact Array Broadband Backend \citep[CABB;][]{2011MNRAS.416..832W} observation and four pre-CABB observations (CX082, C514, C287 and C389). Projects C514 and C287 are made up of numerous pointings, whereas the others are single pointing observations. The projects obtained from the ATCA pre-CABB epoch have central frequencies of 1.38~GHz, 2.37~GHz, 4.79~GHz and 8.64~GHz, and bandwidth of 128~MHz with 32 channels each. The 2.1~GHz CABB dataset have 2~GHz bandwidth, resulting in significantly better sensitivity. The $(u,v)$ data for each frequency band and epoch are automatically flagged for outliers using the {\tt pgflag} \citep{2010MNRAS.405..155O} task and calibrated using standard calibration procedures in the {\sc MIRIAD} \citep{1995ASPC...77..433S} package. The missing short spacings from all these ATCA radio images impair detection of large scale structure in the galaxy\footnote{The shortest baseline of the 375-m array at $\lambda$=6 and 3~cm is 31~m, which is sensitive to structure up to sizes of 6\arcmin.}.

A natural weighting scheme is used to produce an image with the lowest r.m.s. while sacrificing angular resolution. A second set of images is also created with a robust parameter of $-2$, corresponding to uniform weighting. However, the higher angular resolution still does not resolve many sources and therefore gives very little additional morphological information. As suggested by \citet{1998A&AS..130..421F}, variable sources could exhibit different flux densities at different epochs (as well as at different frequencies) and therefore could lead to misleading spectral index estimates. Based on a number of radio catalogues that we produced in other nearby galaxies, we expect that not many of our sources may be affected by this process. 

For the projects C514/C389 and CX082/C287 with numerous pointings, joint deconvolution was implemented using the {\sc MIRIAD} {\tt clean} task. Images were made individually for each project to ensure adequate flagging and calibration. Then for data-sets with the same frequency, joint deconvolution was implemented to combine images. Images from the pre-CABB projects are also made without the sixth antenna as they had a sparse $(u,v)$ coverage. The details for the resulting images from these data-sets are provided at the bottom part of Table~\ref{images}.

\subsubsection{SN2005at in ATCA Radio Images}
 \label{sn}

\citet{2014A&A...572A..75K} reported detecting SN2005at at $\sim9\sigma$ in their radio observations (ATCA project CX082, observed 30$^{th}$ March 2005 for $\sim$2h) at 1.38 and 2.37~GHz, with 4 active antennae yielding 6 baselines. We reconstruct images from these particular observations, obtained from the ATOA archive, following \citet{2014A&A...572A..75K}. We use the same 4 antennas (CA01, CA02, CA05 and CA06) with a robust parameter of 1.5. The reconstructed images are of poor quality due to a lack of $(u,v)$ coverage and have a similar noise level (0.15--0.17~mJy~beam$^{-1}$) as in \citet{2014A&A...572A..75K}. Although the peak flux density estimated from our images at the location of SN2005at is slightly above 5$\sigma$ (0.826~mJy~beam$^{-1}$) at 1.38~GHz (but 3.9$\sigma$, 0.658~mJy~beam$^{-1}$ at 2.37~GHz), we have little confidence in the detection of SN2005at due to the very poor image quality where a lot of the sidelobe and artefact patterns are obvious. This source does not appear in our other images. Also, we did combine the data from CX082 with C287 which was observed several years before the SN event, and this could cause any detection in CX082 to be averaged down below the noise level. We do not use the CX082 observations (alone) any further in this study. 

The radio brightness of SNe such as SN2005at (type~Ic; stripped envelope SNe) would be expected to increase only in the first few hundred days since its discovery \citep{2014MNRAS.440.1067R}. 
However, while somewhat untypical and different to SN2005at, SN1987A \citep{2013ApJ...777..131N} or the very young SNR G1.9+1.3 \citep{2014SerAJ.189...41D} are still increasing its radio flux density. In subsequent, more sensitive ATCA-CABB observations, SN2005at was not detected. 

\subsubsection{GLEAM Data}

\begin{figure*}
 \begin{center}
  \includegraphics[trim=0 0 0 0, width=0.95\linewidth]{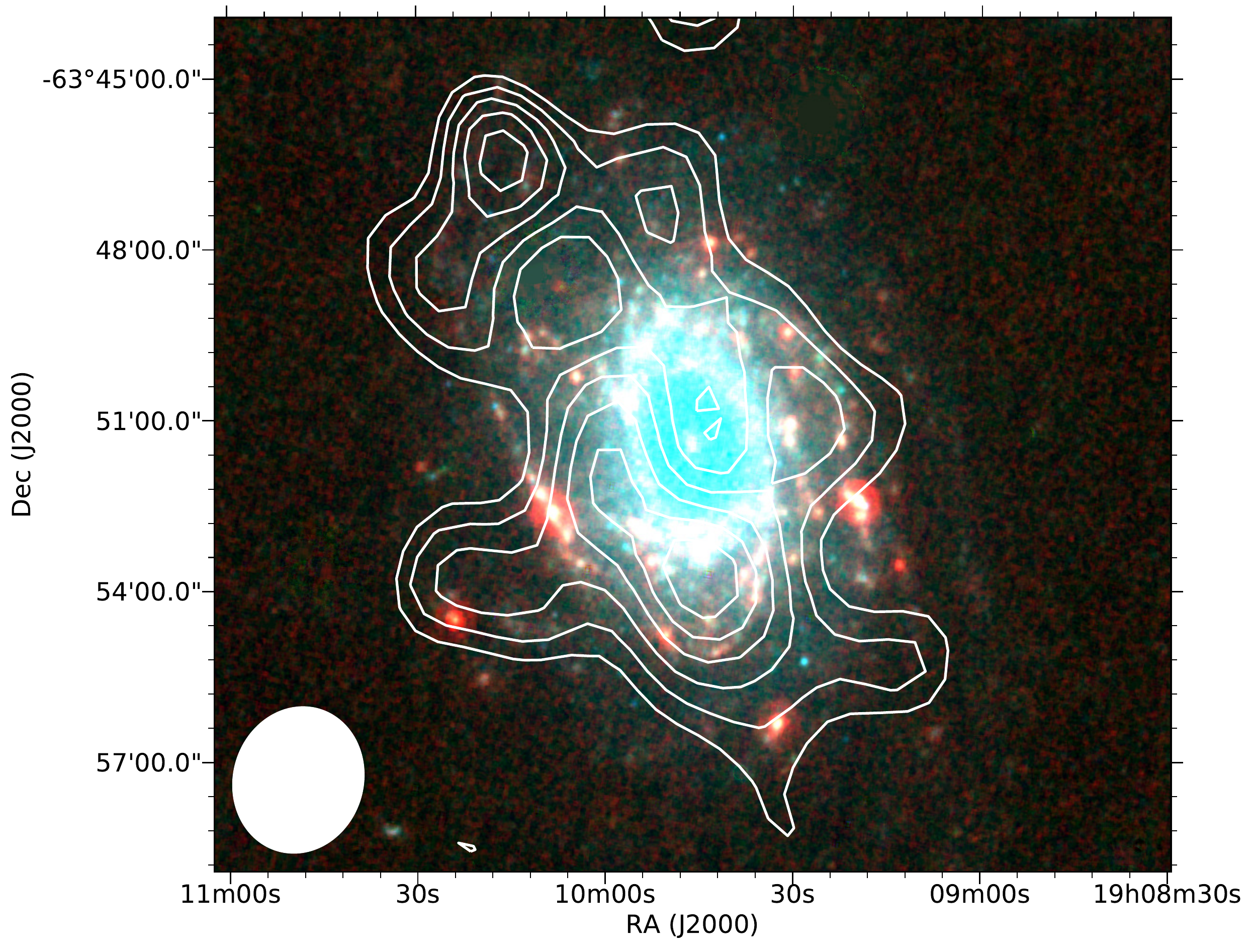}
  \caption{WISE RGB bands W4+W2+W1 image overlaid with MWA 200~MHz image contours. The MWA contours (white) are 4$\sigma$ to 10$\sigma$ with spacings of 1$\sigma$ and the 200~MHz beam is shown in the lower left.}
  \label{wisemwa}
 \end{center}
\end{figure*}

We also examine radio continuum data from the Galactic and Extragalactic All-Sky Murchison Widefield Array \citep[GLEAM;][]{2013PASA...30....7T,2015PASA...32...25W} in our analysis. The GLEAM survey provides low-frequency (72--231~MHz) low angular resolution (5\arcmin --1.7\arcmin) radio images with excellent surface brightness and large-angular-scale sensitivity\footnote{GLEAM's shortest baseline of 7~m gives sensitivity to structures up to 29$^{\circ}$--10$^{\circ}$ at 76--227~MHz respectively, as per Table~4 of \citet{2017arXiv170202434K}}. Here we use Extragalactic GLEAM Catalogue images\footnote{\url{http://mwa-web.icrar.org/gleam\_postage/q/form}} \citep[EGC;][]{2017MNRAS.464.1146H} at four frequency bands, of which three have 30~MHz bandwidth and are centred on 88~MHz, 118~MHz and 154~MHz, and one has 60~MHz bandwidth centred on 200~MHz (see Figure~\ref{wisemwa}). The r.m.s. noise levels of the images are: 19~mJy~beam$^{-1}$ at 200~MHz, 24~mJy~beam$^{-1}$ at 154~MHz, 26~mJy~beam$^{-1}$ at 118~MHz and 57~mJy~beam$^{-1}$ at 88~MHz. The details of calibration and imaging procedure of the GLEAM data are described in \citet{2017MNRAS.464.1146H}.

%
%

\subsection{Infrared Data}

The {\it WISE} telescope surveyed the entire sky at wavelengths of 3.4~$\mu$m, 4.6~$\mu$m, 12~$\mu$m and 22~$\mu$m \citep[{\it WISE} bands W1-W4 respectively;][]{wright2010wide, 2017ApJ...836..182J}, in which the photometric bands were chosen to be sensitive to both stellar light and interstellar medium processes \citep{2011ApJ...735..112J}. The near-IR bands W1 and W2, tracing the evolved stellar population, have been used to measure the underlying stellar mass. The mid-IR molecular-PAH emission W3 band and warm-dust W4 band have been used to measure star-formation activity. The mosaic images presented here are produced using the ICORE software package \citep{2013arXiv1301.2718M} which resamples the stack of raw, single-frame images to 1\arcsec\ pixels. This is done by using a `drizzle' technique, achieving native resolutions of 5.9\arcsec, 6.5\arcsec, 7.0\arcsec\ and 12.4\arcsec\ at 3.4~$\mu$m, 4.6~$\mu$m, 12~$\mu$m and 22~$\mu$m, respectively. This represents a $\sim$30\% improvement from the public release ``Atlas" images which are smoothed for optimal point source detection \citep{2012AJ....144...68J}.

\begin{figure*}
\begin{center}
\includegraphics[width=\linewidth]{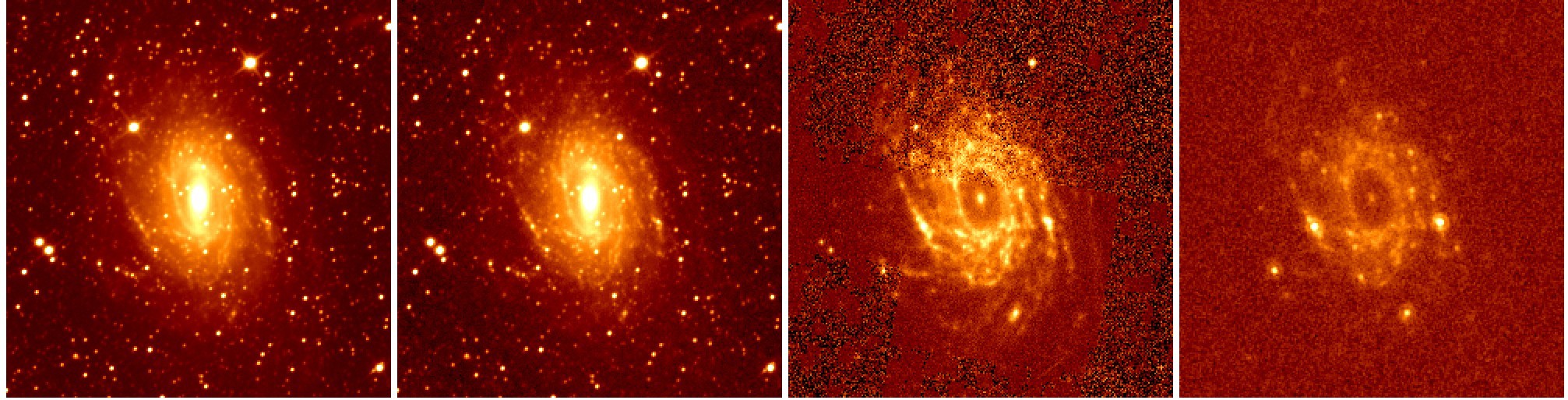}
\caption{{\it WISE} imaging of \ngc. The panels show the four bands of {\it WISE} (left to right: W1, W2, W3 and W4). Each panel has an angular size of 16.6\arcmin.} 
\label{wiseimage}
\end{center}
\end{figure*}

\begin{figure*}
\begin{center}
\includegraphics[width=\linewidth]{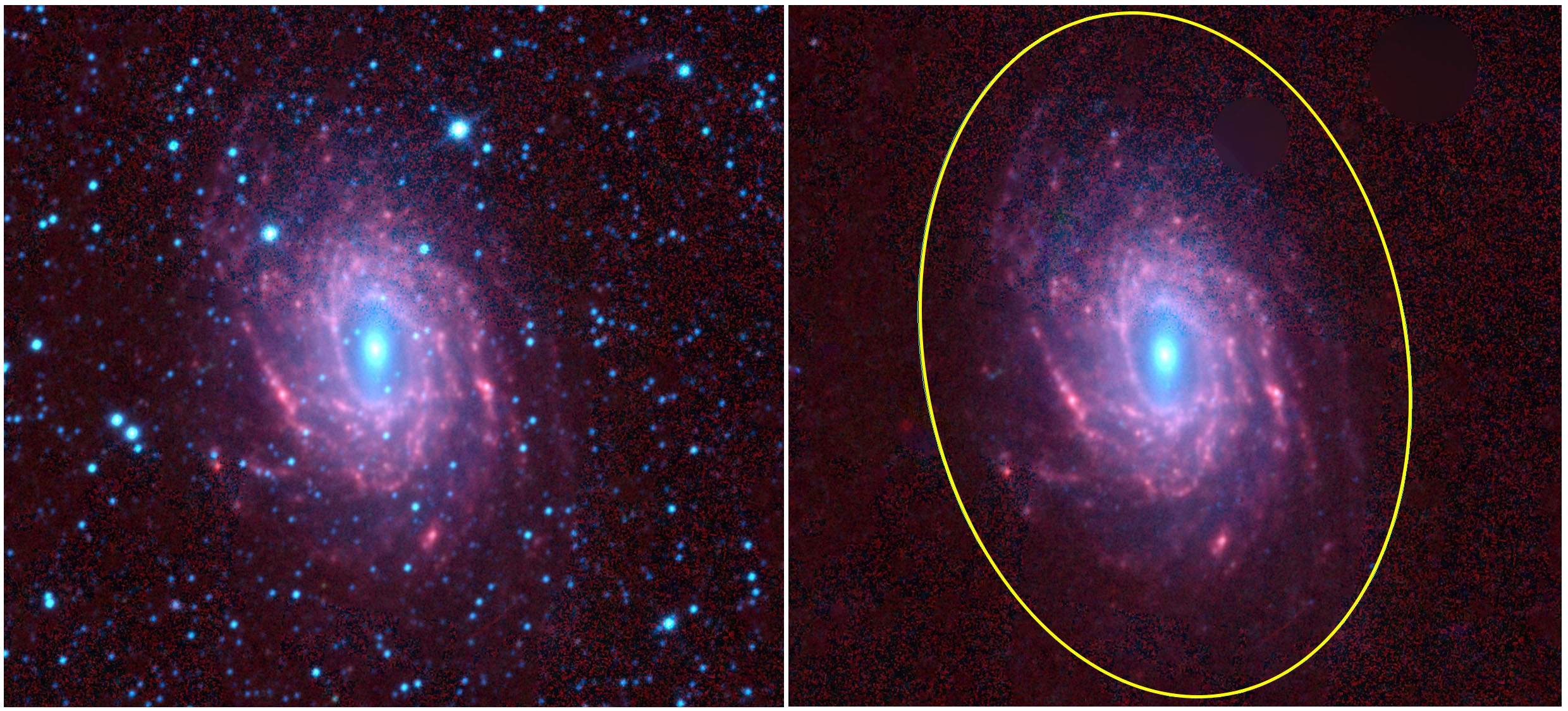}
\caption{{\it WISE} 3-color (W1+W2+W3) image of \ngc.  Left panel shows the images before star subtraction, and the right panel after. The ellipse represents the W1 isophote at the 1$\sigma$ noise level, and has a diameter of 18.3\arcmin\ (where 1\arcsec\ is 46~pc; 1\arcmin\ is 2.76~kpc). 
} 
\label{wiseimage2}
\end{center}
\end{figure*}

Native-resolution mosaic images of \ngc\ in W1--W4 are shown in Figures~\ref{wiseimage} and \ref{wiseimage2}, where the colour image represents a RGB-combined W3, W2 and W1 multi-$\lambda$ window. The evolved stellar population is notably seen in the nucleus and bulge regions (appearing blue/cyan in colour), while the SF regions are concentrated in the arms (appearing orange/red in colour). At the distance of \ngc, the physical pixel scale is 46~pc, while the resolved scale is about 270~pc, which is larger than giant molecular clouds (GMCs), but clearly portraying the star formation complexes where the gas is concentrated.

%

%

\section{RESULTS AND DISCUSSION}
\label{sec:results and discussion}

\subsection{Discrete Sources in \ngc}

\subsubsection{X-Ray Sources}
We detect a total of 117 sources across all four X-ray bands (see Table~\ref{X-ray})\footnote{The complete catalogue is available in the electronic version only.}. We estimate the minimum detectable flux of the X-ray sources to be 8.23$\times10^{-16}$~ergs~s$^{-1}$~cm$^{-2}$ using the broad X-ray band, and the luminosity to be 8.89$\times10^{36}$~ergs~s$^{-1}$. 

Table~\ref{X-ray} lists the positions and intensities of the sources along with the luminosities and hardness ratios. We detect four off-centre sources (J190845-634817, J191104-634611, J191107-634823, J191109-635258) along with the slightly off-centre central source (J191013-635404) which were previously detected by \citet{1992ApJS...80..531F} using {\it Einstein} observations. We also search for foreground sources in the field of \ngc\ by cross-checking the positions of the X-ray catalog with SIMBAD. 
Using a search radius of 12\arcsec\ we found one result --- the K0-type star 190449.9$-$635729 \citep{1984MNRAS.210..373P}.


\begin{table*}[t]

 \small
 \caption{X-Ray source catalogue sample. The luminosity estimates are accurate to $\sim$30\%. The count rate in Column~8 corresponds to the broad band. For sources with 2 intensity values (excluding broad band), we invoked upper limits on the soft, medium and hard bands ($6.070$, $3.200$, and $4.370$~photons~s$^{-1}$~cm$^{-2}$, respectively) for the remaining intensity to get a hardness ratio. $H\!R_1$ is derived from equation~\ref{eq12} with typical errors of 23\% while $H\!R_2$ is derived from equation~\ref{eq11} with errors of 25\%. Source J190946-635129 is the central black hole. The complete catalogue is available in the electronic version only.  \label{X-ray}}
 \begin{center}
\rotatebox{90}{
\begin{tabular}{ccccccccccc}
\hline
  \multicolumn{1}{p{1.5cm}}{\centering X-ray Source \\ Name} &
  \multicolumn{1}{p{1.1cm}}{\centering RA \\ (J2000)  \\ (h m s)}              &
  \multicolumn{1}{p{1.4cm}}{\centering DEC \\ (J2000) \\ ($^\circ$ \arcmin\ \arcsec)} &
  \multicolumn{1}{p{2cm}}{\centering $E_\mathrm{Broad}$ \\ (ergs~s$^{-1}$~cm$^{-2}$)\\ ($\times 10^{-15}$)} &
  \multicolumn{1}{p{2cm}}{\centering $E_\mathrm{Soft}$  \\ (phs~s$^{-1}$~cm$^{-2}$) \\ ($\times 10^{-5}$)} &
  \multicolumn{1}{p{2cm}}{\centering $E_\mathrm{Med}$   \\ (phs~s$^{-1}$~cm$^{-2}$) \\ ($\times 10^{-5}$)} &
  \multicolumn{1}{p{2cm}}{\centering $E_\mathrm{Hard}$  \\ (phs~s$^{-1}$~cm$^{-2}$) \\ ($\times 10^{-5}$)} &
  \multicolumn{1}{p{1cm}}{\centering Count Rate  \\ (cts~s$^{-1}$)} &
  \multicolumn{1}{p{0.6cm}}{\centering L$_\mathrm{Bol}$    \\ (ergs~s$^{-1}$) \\ ($\times 10^{37}$)} &
  \multicolumn{1}{p{0.4cm}}{\centering $H\!R_1$} &
  \multicolumn{1}{p{0.4cm}}{\centering $H\!R_2$} \\
\hline
J190823-633657 & 19:08:23.6 & --63:36:57.98 &  32.40 & 0.126 & 0.281 & 1.230 &  200$\pm$8.0 & 34.9 &    0.38 &	0.63 \\
J190840-633802 & 19:08:40.6 & --63:38:02.32 &  27.20 & 0.497 & 0.196 & 0.429 &  200$\pm$7.9 & 29.4 & --0.43 &      0.37 \\
J190842-635154 & 19:08:42.1 & --63:51:54.91 &   6.52 & $\textless$6.070 &  0.073 &  $\textless$4.370 &  31$\pm$3.1 & 7.04 &  0.99 &  --0.99 \\
J190842-635105 & 19:08:42.2 & --63:51:05.59 &   1.65 &	  & 	  & 	  &  20$\pm$2.5 & 1.78 & 	&	   \\
J190845-634817 & 19:08:45.0 & --63:48:17.92 &  52.70 &   1.040 &     0.417 &   0.624 & 60$\pm$4.3  & 56.9 &  --0.43 &        0.19 \\
\smallskip
J190908-634638 & 19:09:08.5 & --63:46:38.89 & 34.90 & 0.378 & 0.350 & 0.637 &  56$\pm$4.2 & 37.7 & --0.04 & 0.29 \\
J190909-635651 & 19:09:09.3 & --63:56:51.71 &  7.50 & $\textless$6.070 &     0.068 &   0.122 &   27$\pm$2.9 & 8.10 & --0.98 &        0.28 \\
J190910-635328 & 19:09:10.1 & --63:53:28.81 &  4.23 &   0.178 &                  &                & 25$\pm$2.8  & 4.57 &  &                 \\
J190911-635551 & 19:09:11.0 & --63:55:51.11 & 12.00 &   0.325 &     0.109 &   0.200 & 21$\pm$2.5 & 12.9 &  --0.49 &        0.29 \\
J190911-634529 & 19:09:11.9 & --63:45:29.47 &  2.35 & & & 0.283 & 12$\pm$1.9 & 2.54 & & \\
\smallskip
J190916-634033 & 19:09:16.6 & --63:40:33.47 & 13.00 & 0.261 & 0.163 & $\textless$4.370 & 28$\pm$3.0 & 14.0 & --0.23 & 0.93 \\
J190918-634552 & 19:09:18.3 & --63:45:52.56 &	 & & 0.036 & & 9.1$\pm$1.7 & & & \\
J190925-634331 & 19:09:25.5 & --63:43:31.51 &  7.87 & 		 $\textless$6.070 & 0.122 & 0.174 & 35$\pm$3.3 & 8.50 & --0.96 & 0.18 \\
J191025-634506 & 19:10:25.0 & --63:45:06.32 & 4.25 &  &  &  & 13$\pm$2.1  & 4.59 & &  \\
J191035-635008 & 19:10:35.0 & --63:50:08.55 & 5.52 & 0.129 & & 0.199 &  29$\pm$3.0   & 5.96 & & \\
\smallskip
J191042-634553 & 19:10:42.9 & --63:45:53.08 & 13.10 & $\textless$6.070 & 0.146 & 0.255 & 29$\pm$3.0  & 14.2 & --0.95 & 0.27 \\
J191045-635240 & 19:10:45.1 & --63:52:40.64 & 4.58 &  &  & 0.133 & 20$\pm$2.5  & 4.94 &  & \\
J191056-634548 & 19:10:56.8 & --63:45:48.98 & 69.90 & 0.934 & 0.518 & 0.817 & 88$\pm$5.2  & 75.5 & --0.29 & 0.22 \\
J191104-634611 & 19:11:04.8 & --63:46:11.50 & 	 & & & 0.232 & 36$\pm$3.3 & & \\
J191107-634823 & 19:11:07.3 & --63:48:23.50 & 2.11 & 0.435 & 0.156 & 0.396 & 30$\pm$3.1  & 2.28 & --0.47 & 0.44 \\
\hline
\end{tabular}
}
\end{center}

\end{table*}

\subsubsection{Radio-continuum Sources}

In Table~\ref{commonspec}, we present a sample of our combined radio catalogue\footnote{The complete catalogue is available in the electronic version of the paper only.}. We extract radio sources from each image using the source extraction package Python Blob Detection and Source Measurement \citep[PyBDSM;][]{mohan2015pybdsm} and use local r.m.s values that vary across the image. For all detected sources, we list the position and integrated flux density at each given frequency. We estimate that the uncertainty in flux density measurements is below 10\%. Our most sensitive and among the highest-resolution radio-continuum images is the 2.1~GHz ($\lambda$=13~cm) C2697 CABB data-set (r.m.s of 0.010~mJy~beam$^{-1}$ and FWHM of 7.51\arcsec$\times$4.41\arcsec). From this image, a total of 280 unique radio sources are identified, and catalogued in Table~\ref{commonspec}. However, SN2005at, at RA~(J2000)=19$^h$09$^m$53.6$^s$, Dec~(J2000)=--63\D49\arcmin24.1\arcsec\  \citep{2005CBET..119....1M} could not be detected in any other of our radio images down to a 2$\sigma$ level (see Section~\ref{sn}).

\subsubsection{Source Classification}
 \label{crit}

Radio continuum sources found towards \ngc\ are likely to fall into three classes: \HII\ regions and SNRs within \ngc; and background radio sources. We initially classify our sources into these categories based on simple criteria:
\begin{enumerate}
\item Sources detected in both radio and IR are taken to be \HII\ regions;
\item Sources detected in both radio and X-rays are taken to be SNRs;
\item Sources outside the optical extent of \ngc\ are taken to be exclusively background sources. However, we also expect that some of the sources listed here might be associated with \ngc, while some that are within the defined area of \ngc\ could be of a background nature.
\end{enumerate}


Following \citet{2013A&A...558A.101S} cross-matching the SMC radio and the X-ray data, we use the most accurate position based on the PSF maps of the X-ray images. We set a search radius of 10\arcsec\ between the X-ray and all radio positions in order to match the radio catalogues from different frequencies to each other. The only exception was matching the 2.1~GHz data to the 1.4~GHz pre-CABB data, for which we use 30\arcsec, due to the poor resolution of the 1.4~GHz image. From 117 X-ray and 280 radio catalogued sources, we find only nine sources detected in both radio and X-rays. Namely, they are: J191042-634553, J191107-634826, J190916-634033, J190925-634329, J190908-634638, J190946-635128, J190935-635241, J190919-635329 and J190840-633810. Three of these nine sources are outside the optical extent of \ngc\ while six are considered as SNRs and SNR candidates. Moreover, we find 44 sources in common to radio and IR surveys. Depending on their characteristics and position relative to the \ngc\ we classify them as either \HII\ regions (17), background sources (25) or in two other cases as central black hole (Section~\ref{agn}) and SNR (J190908-634638; see Section~\ref{snrssec}).


We estimate radio spectral indices for 10 sources detected in more than two radio bands and also in X-ray. We emphasise that the radio images used in estimating spectral index are not of the same resolution, and as such, for extended sources, the varying angular resolution may cause the spectrum to be less smooth than the true one or steeper at higher frequencies. It is well established that the radio spectral index (where \mbox{$S\propto \nu^{\alpha}$}) overlaps between different groups of sources and as such cannot be exclusively used for source classification. However, radio sources with steep spectral index of $\alpha< -0.9$  are more likely to be background sources and those with flatter spectral indices ($\alpha< -0.5$) slightly favour an \HII\ classification. In between two, SNRs tend to have $\alpha=-0.5\pm0.3$. However, background sources which we expect to be a dominant population in our catalogues exhibit a wider range of spectral indices. We do not expect to find any pulsars (which are usually steep spectrum sources) and Pulsar Wind Nebulae (flat spectrum) in \ngc\ as they would emit at levels below the detection limit based on our observations from other nearby galaxies where no such sources could be detected \citep[e.g.][]{Brien2013}. 


We visually inspected and classified the sources in the radio, X-ray and IR images and classified sources according to their morphology from the best available resolution band, spectral index and hardness ratio similar to \citet{1998A&AS..130..421F}. 


Using these criteria, we classify all sources found in this study into these three groups: there are 254 background sources of which 229 are outside of the galaxy (marked in Table~\ref{commonspec} as ``BKG$\star$"), 14 are likely background objects in the vicinity of \ngc\ (marked in Table~\ref{commonspec} as ``BKG"), and additional 11 are classified as likely background objects (marked in Table~\ref{commonspec} as ``bkg"). We also find 2 highly probable SNRs and 5 SNR candidates based on the strength of X-ray emission (marked in Table~\ref{commonspec} as ``snr"), discussed further in Section~\ref{snrssec}. Finally, we classify 17 discrete sources as likely \HII\ regions (Section~\ref{h2reg}) and one as AGN (central \ngc\ black hole; Section~\ref{agn}). Radio source J190953-635059 is the only source with an inverted spectrum of $\alpha$=0.32, and it has no X-ray or IR detection. We tentatively classify this source as a background.


\begin{table*}[p]
 \small
 \hspace*{-2cm}
 \caption{A sample list of radio point sources in the \ngc\ field at $\lambda$=20~cm, 13~cm, 6~cm and 3~cm. Column~8 is the spectral index for all flux density measurements of a source that had three or more flux density values. The last column is based on the classification scheme outlined in Section~\ref{crit}. Sources which are strong candidates for their respective classification are presented in upper case, while less robust classifications are presented in lower case. Sources marked with an asterisk are outside of the galaxy ellipse defined as 20\arcmin$\times$12.9\arcmin. The estimated uncertainty in flux density is below 10\%.}
 \begin{center}
\begin{tabular}{ccccccccccl}
\hline
  \multicolumn{1}{p{1.8cm}}{\centering Radio Source \\ Name} &
  \multicolumn{1}{p{1.2cm}}{\centering RA~(J2000)  \\ (h m s)}              &
  \multicolumn{1}{p{1.5cm}}{\centering DEC~(J2000) \\ (\D\ \arcmin\ \arcsec)} &
  \multicolumn{1}{p{0.8cm}}{\centering S$_\mathrm{20cm}$   \\ (mJy)}                &
  \multicolumn{1}{p{0.8cm}}{\centering S$_\mathrm{13cm}$   \\ (mJy)}                &
  \multicolumn{1}{p{0.8cm}}{\centering S$_\mathrm{6cm}$    \\ (mJy)}                &
  \multicolumn{1}{p{0.8cm}}{\centering S$_\mathrm{3cm}$    \\ (mJy)}                &
  \multicolumn{1}{p{1.2cm}}{\centering $\alpha \pm \Delta \alpha$}         &
  \multicolumn{1}{p{0.5cm}}{\centering IR           \\ ID}                   &
  \multicolumn{1}{p{1cm}}{\centering X-Ray        \\ ID}                   &
  \multicolumn{1}{p{0.6cm}}{\centering Type}                 \\
\hline
J190908-634638 & 19:09:08.5 & --63:46:38.52 &	 & 0.089 &       &       &                 & Y & Y & SNR \\
J190916-634033 & 19:09:16.8 & --63:40:33.79 & 3.130 & 0.791 &       &       &                 &   & Y & SNR \\
J190919-635241 & 19:09:19.3 & --63:52:41.70 & 1.610 & 0.312 &       &       &                 &   &   & bkg \\
J190919-635329 & 19:09:19.8 & --63:53:29.73 & 	 & 0.196 &       &       &                 &   & Y & snr\\
\smallskip
J190920-635223 & 19:09:20.0 & --63:52:23.90 & 1.610 & 1.640 & 0.929 & 0.843 & --0.54$\pm$0.12 & Y &   & \HII \\
J190932-635619 & 19:09:32.4 & --63:56:19.62 & 0.500 & 0.450 & 0.390 &       & --0.20$\pm$0.02 & Y &   & \HII \\
J190942-635542 & 19:09:42.4 & --63:55:42.05 & 3.370 & 3.040 & 1.072 &       & --0.91$\pm$0.41 &   &   & bkg \\
J190943-634754 & 19:09:43.1 & --63:47:54.43 &	 & 0.243 &       &       &                 & Y &   & BKG \\
J190943-635429 & 19:09:43.7 & --63:54:29.93 & 1.050 & 0.663 &       &       &                 & Y &   & \HII \\
\smallskip
J190946-635128 & 19:09:46.3 & --63:51:28.00 &  & 0.126 &       &       &                 & Y & Y & AGN \\
J190953-635059 & 19:09:53.1 & --63:50:59.31 & 1.500 & 2.100 & 2.960 & 2.690 & 0.32$\pm$0.12   &   &   & BKG \\
J191007-634708 & 19:10:07.6 & --63:47:08.64 & 1.730 & 1.70 & 0.823 &       & --0.69$\pm$0.33 &   &   & bkg \\
J191008-635238 & 19:10:08.4 & --63:52:38.16 & 1.580 & 1.180 & 0.666 &       & --0.70$\pm$0.02 & Y &   & \HII \\
J191010-635218 & 19:10:10.6 & --63:52:18.35 & 1.580 & 0.400 &       &       &                 & Y &   & \HII \\
\smallskip
J191020-634633 & 19:10:20.0 & --63:46:33.51 & 9.250 & 7.440 & 2.472 &       & --0.94$\pm$0.36 & Y &   & bkg \\
J191022-635155 & 19:10:22.0 & --63:51:55.01 &  & 0.768 & 0.639 & 0.811 & --0.05$\pm$0.14 &   &   & BKG$\star$ \\
J191035-634744 & 19:10:35.8 & --63:47:44.03 & 2.680 & 2.510 & 0.793 &       & --0.77$\pm$0.52 &   &   & BKG$\star$ \\
J191042-634553 & 19:10:42.5 & --63:45:53.26 &  & 0.064 &	    &	    &		      &   & Y & snr \\
J191048-635134 & 19:10:48.8 & --63:51:34.94 & 4.220 & 3.620 & 1.050 &	    & --0.72$\pm$0.45 &   &   & BKG* \\
J191146-634902 & 19:11:46.0 & --63:49:02.72 & 2.790 & 1.030 &       &       &                 & Y &   & BKG$\star$ \\
\hline
\end{tabular}
 \label{commonspec}
 \end{center}
\end{table*}

\subsubsection{Central Black Hole (AGN) Source}
 \label{agn}

\begin{figure}
 \begin{center}
  \includegraphics[trim=0 0 0 0, width=1\linewidth]{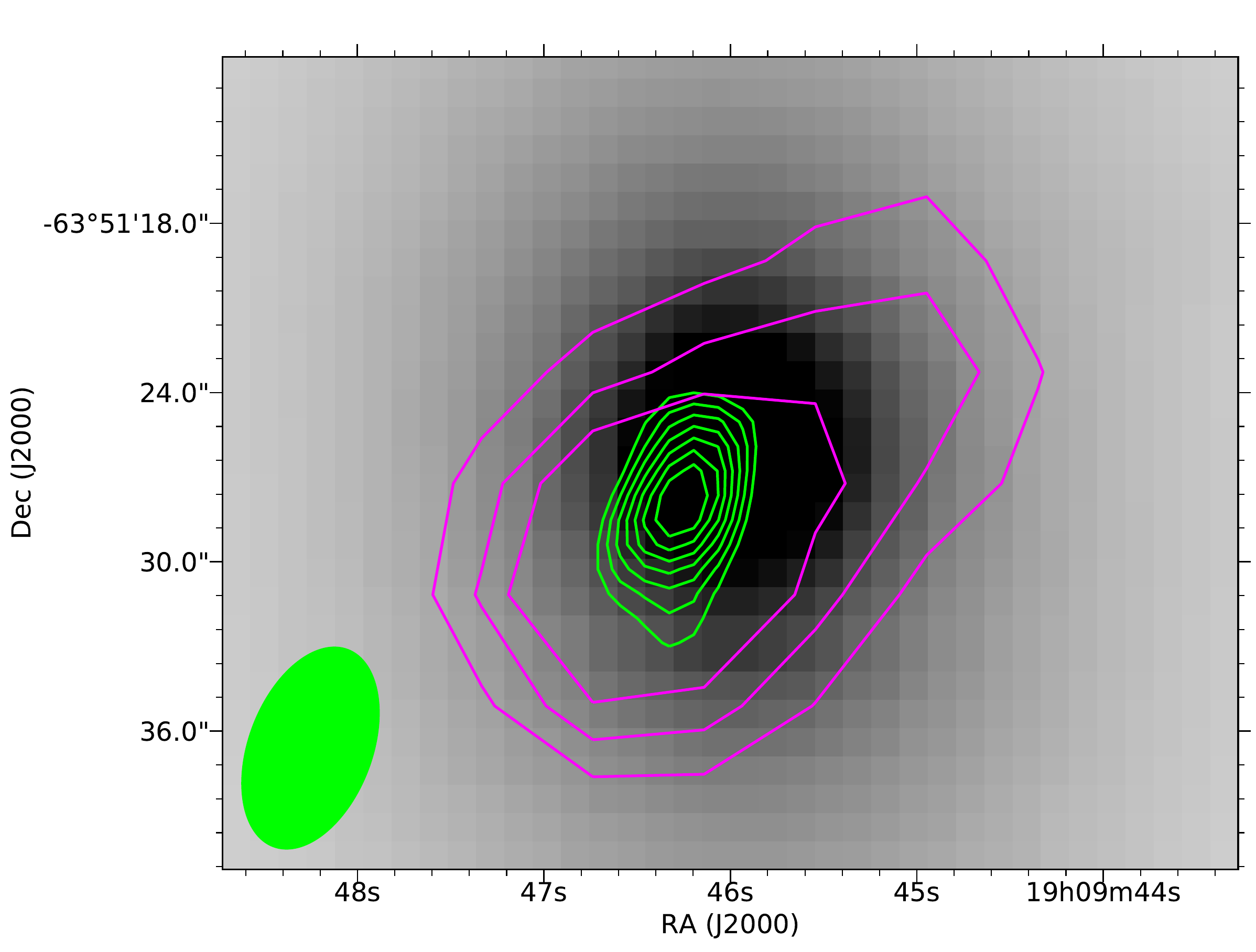}
  \caption{\ngc\ central source J190946-635128. {\it WISE} band 1 image (grey scale) overlaid with radio ATCA 2.1~GHz (project C2697; green contours from 5$\sigma$ to 11$\sigma$ with increments of 1$\sigma$; $\sigma$=0.01~mJy~beam$^{-1}$) and X-ray \chandra\ broad-band (magenta; ID~15384; contours are 1.5$\times 10^{-7}$~cts~s$^{-1}$ to 2.5$\times 10^{-7}$~cts~s$^{-1}$ with increments of 5$\times 10^{-8}$~cts~s$^{-1}$). The beam size is shown in the lower left corner.} 
  \label{blackhole}
 \end{center}
\end{figure}

We identify a central radio and X-ray source as detected by \einstein\ at RA~(J2000)=19$^h$09$^m$46.3$^s$, Dec~(J2000)=--63\D51\arcmin28.00\arcsec\ (source J190946-635128 in Table~\ref{commonspec}) in our 2.1~GHz radio-continuum image as well as in X-ray images (source J190946-635129 in Table~\ref{X-ray}). \citet{1986A&AS...66..335V} listed \ngc\ as hosting an AGN using optical spectroscopy\footnote{Classified as a low-ionization nuclear emission-line region (LINER) in NED database {ned.ipac.caltech.edu}.}. The similarity of \ngc\ to the Milky Way suggests that this source (see Figure~\ref{blackhole}) might be similar to the one in the centre of our Galaxy --- Sagittarius~A$^{\star}$ \citep[Sgr~A$^{\star}$;][]{1997MNRAS.291..219G}

We measure the total flux density of the proposed \ngc\ AGN at 2.1~GHz to be 0.126~mJy. At this flux density level, and assuming a flat spectral index \citep{2013CQGra..30x4003F}, other ATCA observations (such as C389 or CX082) should also detect this central source. However, no other radio continuum detection of this source could be found, probably from a combination of reasons which include (but are not limited to) smaller bandwidth, poorer $(u,v)$ coverage, steeper spectral index or intrinsic source variability. 

Assuming that the spectral index is similar to Sgr~A$^{\star}$ \citep[$\alpha\sim0$]{1998ApJ...499..731F,2013CQGra..30x4003F},\footnote{We detect the central black hole only at one radio frequency (2.1~GHz) and therefore we cannot estimate spectral index.} we estimate that the total bolometric luminosity over the radio spectrum (10~MHz -- 100~GHz) is $\sim1.3\times10^{36}$~ergs~s$^{-1}$\footnote{If $\alpha\sim+0.3$ is assumed then $\nu L_{\nu} \sim3.2\times10^{36}$~ergs~s$^{-1}$.}. This would be an order of magnitude larger than our Milky Way's central supermassive black hole for which the bolometric radio luminosity is estimated to be $\sim$10$^{35}$~ergs~s$^{-1}$ \citep{2013CQGra..30x4003F}. 

The X-ray counterpart of the nucleus did not have sufficient counts for a spectral analysis. The X-ray luminosity of \ngc's supermassive black hole (9.26$\times10^{37}$~ergs~s$^{-1}$) is orders of magnitude larger than that of Sgr~A* \citep[2.0$\times10^{33}$~ergs~s$^{-1}$;][]{2001Natur.413...45B} but comparable to NGC~821 \citep[6$\times10^{38}$~ergs~s$^{-1}$;][]{2007ApJ...667..731P}. This is not surprising as Sgr A* at both wavebands has long been known to be underluminous, with luminosity several orders of magnitude lower than the Eddington luminosity \citep{2003ApJ...591..891B,2013CQGra..30x4003F}. In X-ray images this source is slightly extended along the NW-SE line suggesting possible jet outflows. 
However, our \chandra\ images are sensitive to about 1\arcsec\ (46~pc at the distance of \ngc) which implies that we cannot rule out an association with the other nearby and X-ray luminous sources such as X-ray binaries. 

%
%

\subsubsection{SNRs}
\label{snrssec}

\begin{figure}
 \begin{center}
  \includegraphics[trim=0 10 0 0, width=1\linewidth]{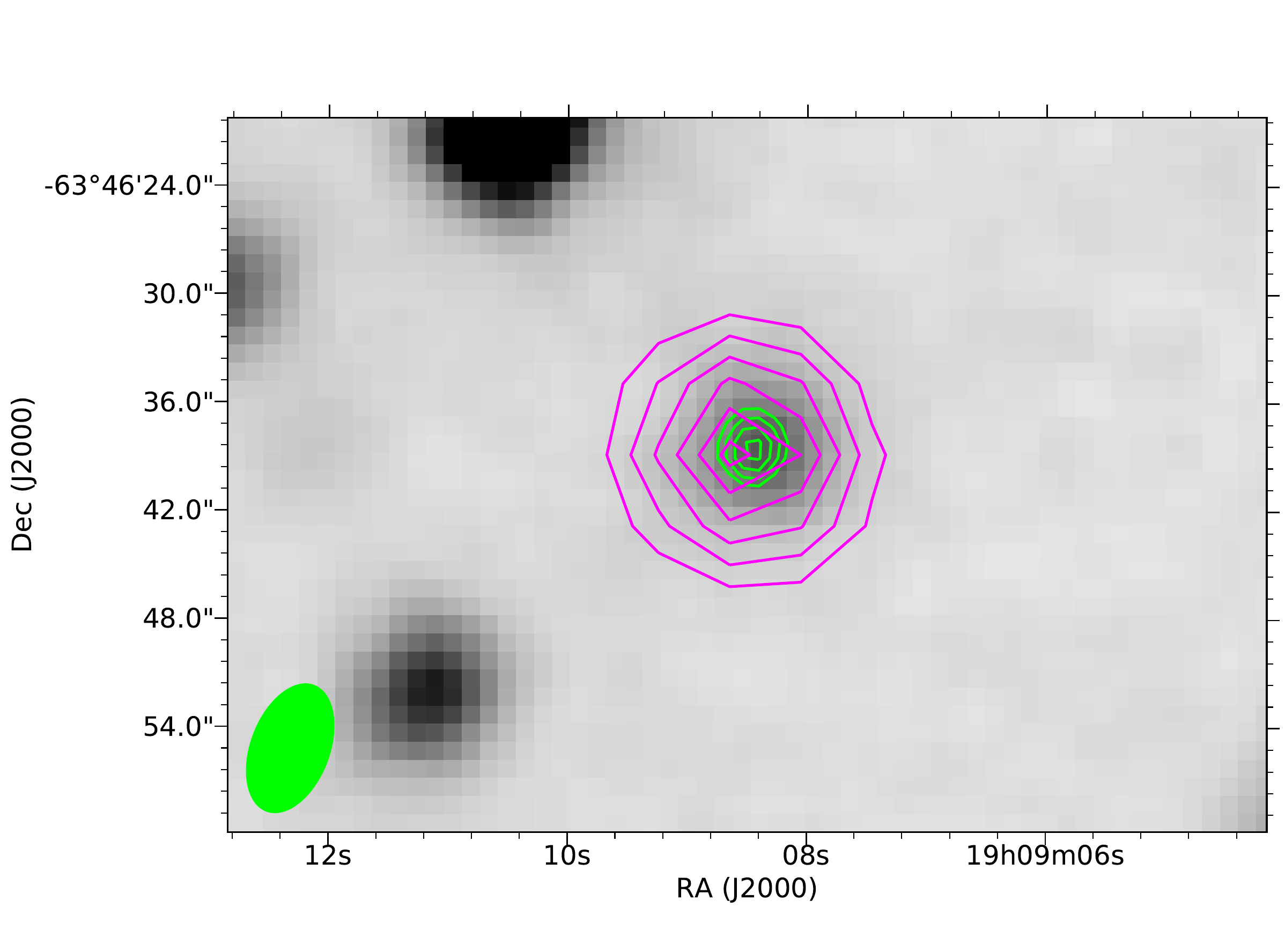}
  \includegraphics[trim=0 0 0 0, width=1\linewidth]{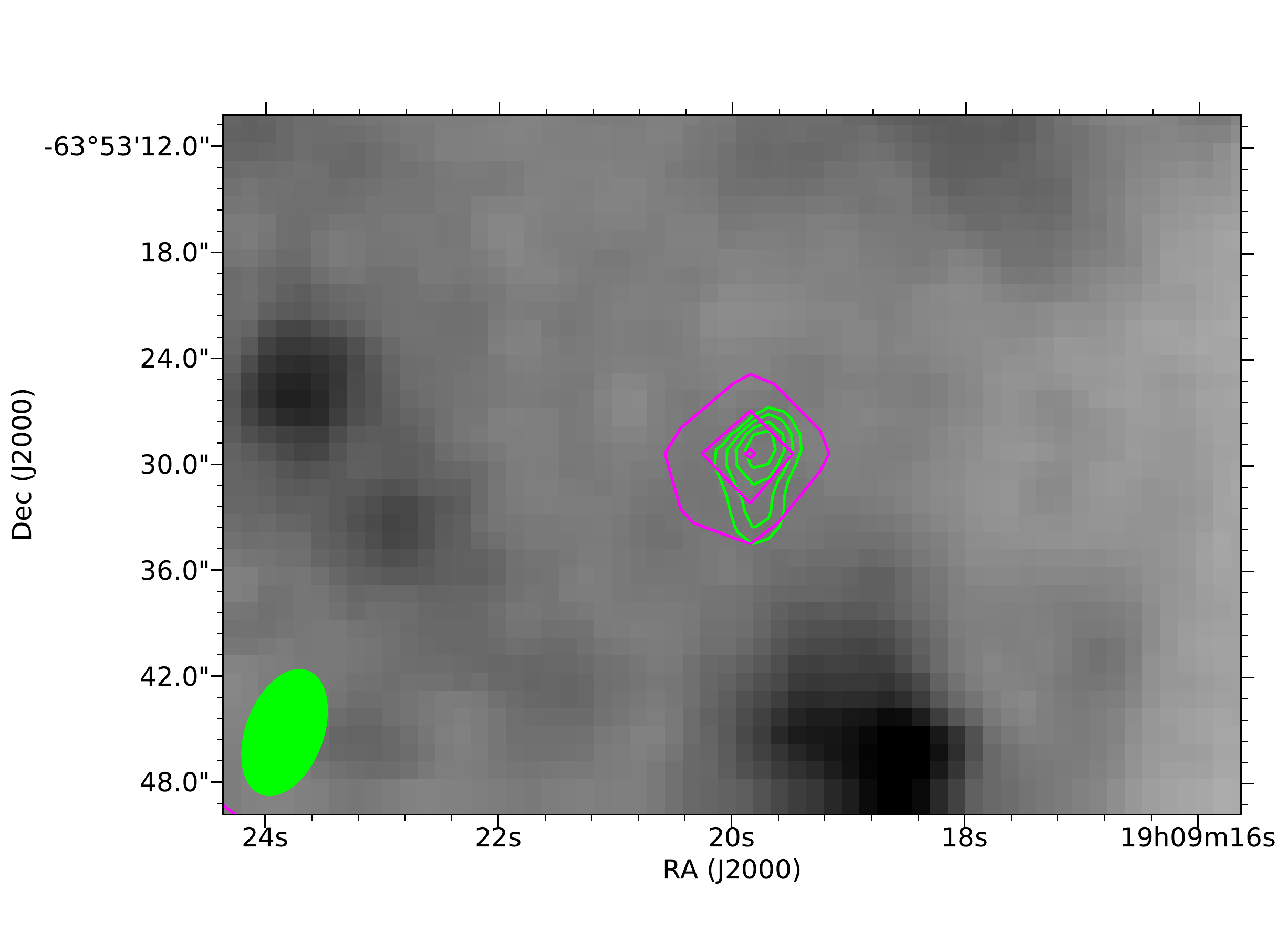}
\caption{SNRs in \ngc. A {\it WISE} band 1 image overlaid with radio ATCA 2.1~GHz (green) and X-ray \chandra\ broad-band (magenta). 
{\it Top}: SNR~J190908-634638 with radio contours from 5$\sigma$ to 8$\sigma$ and spacings 1$\sigma$. The X-ray contours are from 5$\times 10^{-7}$~cts~s$^{-1}$ to 1.75$\times 10^{-6}$~cts~s$^{-1}$ with spacings 2.5$\times 10^{-7}$~cts~s$^{-1}$. 
{\it Bottom}: SNR~J190919-635329 with radio contours from 5$\sigma$ to 8$\sigma$ with spacings 1$\sigma$. The X-ray contours are from 1.5$\times 10^{-7}$~cts~s$^{-1}$ to 2.5$\times 10^{-7}$~cts~s$^{-1}$ with spacings 5$\times 10^{-8}$~cts~s$^{-1}$. The beam is shown in the lower left corner of each image.} 
\label{snr1}
\end{center}
\end{figure}

We use the criteria established in Section~\ref{crit} and classify 7 objects as \ngc\ SNR candidates. We also use {\it WISE} IR images to search for sources with both radio and X-ray emission within the boundaries of \ngc. We add this as a new criterion as we expect that these radio-X-ray-IR detections are most likely to be intrinsic and non-thermal sources i.e. SNRs.  We find only 2 such sources which we classify as the most likely SNRs (see Figure~\ref{snr1}). One of these two sources (J190908-634638) is a very luminous X-ray source -- comparable to young Galactic or Magellanic Clouds SNRs \citep{2015ApJ...803..106R,2016A&A...585A.162M}. However, if this source is a young energetic SNR then its size would not be larger than a few parsec. The other source (SNR~J190919-635329; Figure~\ref{snr1} ({\it bottom})) shows only very weak WISE (IR) emission which is further hampered by the local confusion.

There are also additional 5 sources that are of lower confidence in terms of the SNR classification as they are close to the galaxy boundary . With each source's X-ray to radio detection, we consider these objects more likely to be SNR candidates than background sources. All seven SNRs and SNR candidates are listed in Table~\ref{snrs}.

\begin{table*}[h!]
 \small
 \hspace*{-2cm}
  \caption{Parameters of the two likely SNRs and five SNR candidates in \ngc. The first two listed sources are objects with a high SNR classification confidence, while the last five are SNRs with lower confidence due to being close to the galaxy boundaries. The associated error in the diameter is $\sim$15~pc (2.1~GHz image were deconvolved to allow for the beam size). The surface brightness is estimated using typical SNR spectral index of $\alpha=-0.55$ \citep{1998A&AS..130..421F,2017ApJS..230....2B}.}
 \begin{center}
\begin{tabular}{lcccccccc}
\hline
  \multicolumn{1}{p{2.2cm}}{\centering SNR \ngc \\ Radio Source Name }&
  \multicolumn{1}{p{1.2cm}}{\centering RA \\(J2000)  \\ (h m s)              } &
  \multicolumn{1}{p{1.5cm}}{\centering DEC \\ (J2000) \\ (\D \arcmin \arcsec)  } &
  \multicolumn{1}{p{0.4cm}}{\centering D \\ (pc) \\ }&
  \multicolumn{1}{p{0.5cm}}{\centering S$_{2.1 \rm {GHz}}$ \\ (mJy)  \\ } &
  \multicolumn{1}{p{1.3cm}}{\centering L$_\mathrm{bol_X}$    \\ (ergs~s$^{-1}$) \\ ($\times 10^{37}$)} &
  \multicolumn{1}{p{2.4cm}}{\centering $\Sigma_{1~\rm {GHz}}$ \\ (W~m$^{-2}$~Hz$^{-1}$~sr$^{-1}$) \\ ($\times 10^{-19}$)} &
  \multicolumn{1}{p{2.4cm}}{\centering $\Sigma_{5~\rm {GHz}}$ \\ (W~m$^{-2}$~Hz$^{-1}$~sr$^{-1}$) \\ ($\times 10^{-20}$) } &
  \multicolumn{1}{p{1.5cm}}{\centering L$_{\rm{bol_R}}$ \\ (W~Hz$^{-1}$) \\ ($\times 10^{25}$ )} \\
\hline
J190908-634638 & 19:09:08.5 & --63:46:38.52 & 96 & 0.089 & 37.70   & 0.187 & 0.773 & \p01.07 \\
J190919-635329 & 19:09:19.8 & --63:53:29.73 & 72 & 0.196 & \p03.06 & 0.733 & 3.025 & \p02.35 \\
\\
J190840-633810 & 19:08:40.0 & --63:38:10.68 & 72 & 0.256 & 29.40   & 0.957 & 3.950 & \p03.06 \\
J190916-634033 & 19:09:16.8 & --63:40:33.79 & 107& 0.791 & 14.00   & 1.320 & 5.430 & \p09.47\\
J190925-634329 & 19:09:25.1 & --63:43:29.92 & 48 & 0.259 & \p08.50 & 2.180 & 8.990 & \p03.10 \\
J191042-634553 & 19:10:42.5 & --63:45:53.26 & 60 & 0.064 & 14.20   & 0.345 & 1.420 & 76.60 \\
J191107-634826 & 19:11:07.1 & --63:48:26.54 & 96 & 0.243 & \p02.28 & 0.511 & 2.110 & \p02.91\\
  \hline
\end{tabular}
\label{snrs}
 \end{center}
\end{table*}
%

%
%

%
%

\subsubsection{\HII\ Regions}
\label{h2reg}

\begin{figure*}[h!]
\vspace*{2.5cm}
\begin{center}
\includegraphics[trim=110 0 0 200, width=0.8\columnwidth]{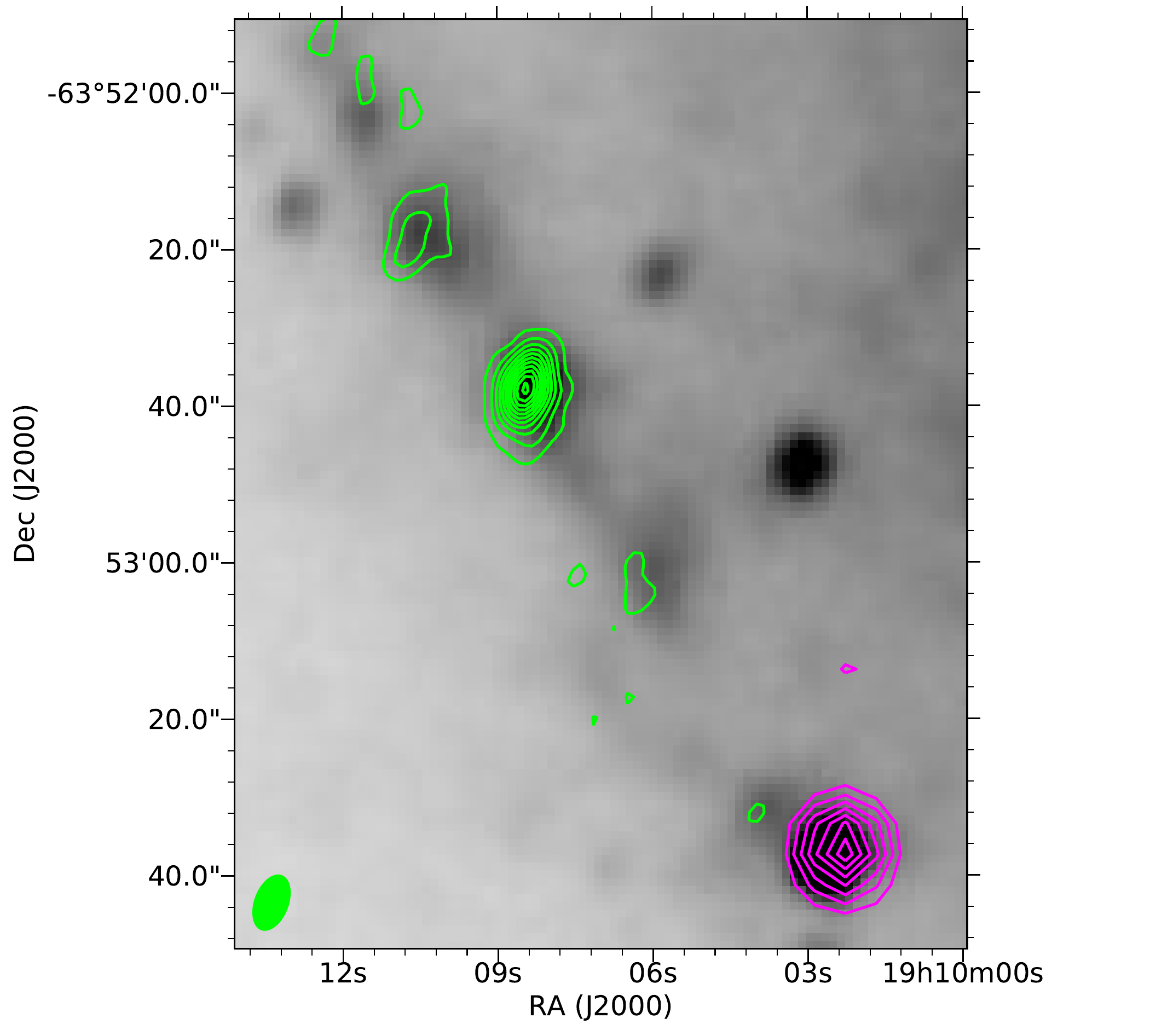}
\includegraphics[trim=65 0 0 200, width=0.8\columnwidth]{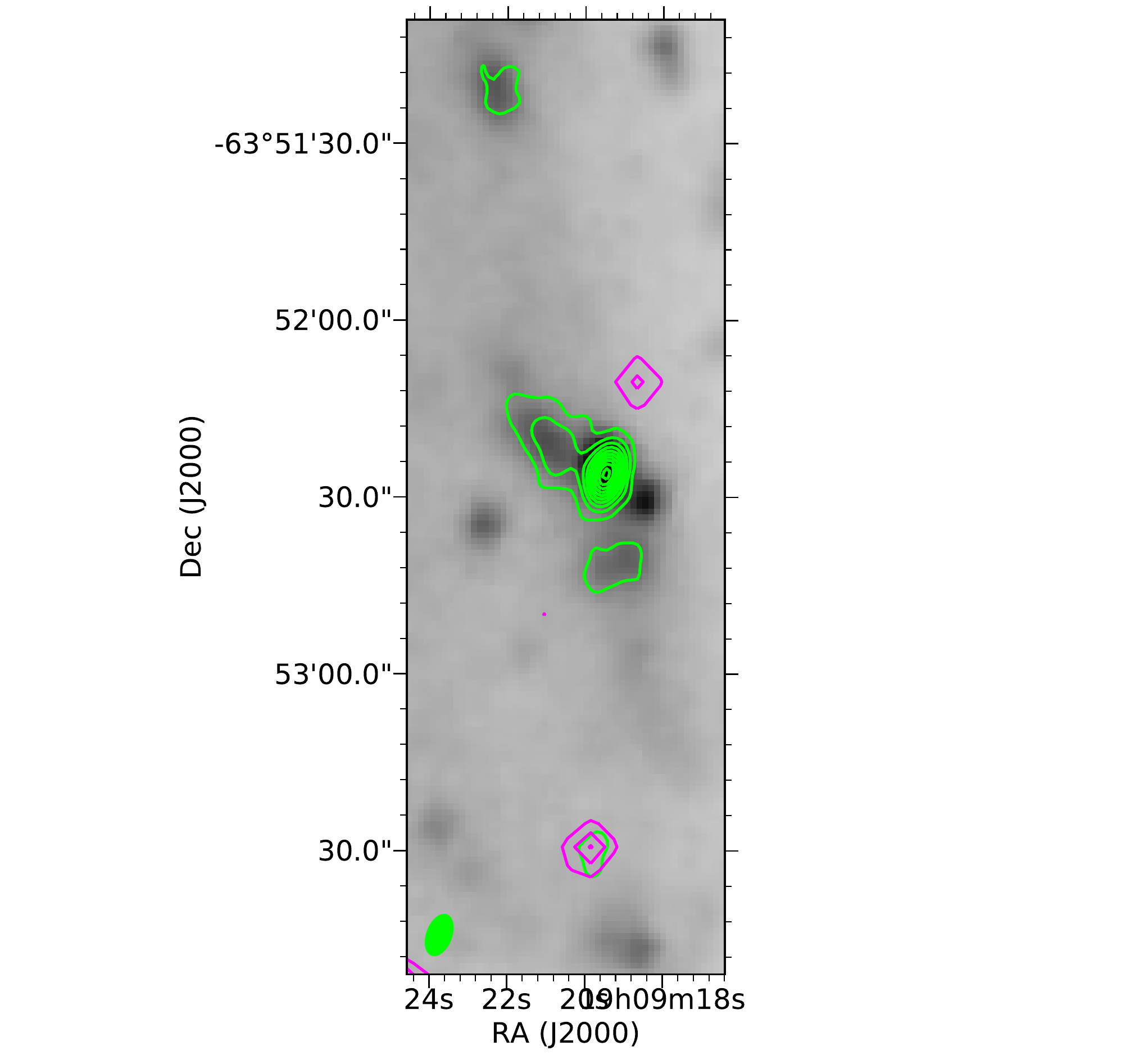}
\caption{Prominent \HII\ regions in the spiral arms of \ngc\ with {\it WISE} band 1 image overlaid with radio ATCA 2.1~GHz (green) (project C2697; 3$\sigma$ to 13$\sigma$ with spacings 1$\sigma$) and X-ray \chandra\ broad-band (magenta) (ID 15384; 1.5$\times 10^{-7}$~cts~s$^{-1}$ to 9$\times 10^{-7}$~cts~s$^{-1}$ with spacings 5$\times 10^{-8}$~cts~s$^{-1}$) contours. The synthesised beam is shown in the lower left corner of each image in green.} 
  \label{starform}
 \end{center}
\end{figure*}

There are 17 radio sources seen in our radio images as concentrations of gas (and stars) in the outer arms of \ngc\ (see Table~\ref{hii}). These sources are also detected as strong point sources in all of the {\it WISE} images. Two of these sources are clearly in the spiral arms of \ngc\ and are shown in Figure~\ref{starform}. However, we do not detect the \HII\ region reported by \citet{1982ApJ...252..594T} at RA~(J2000)=19$^h$10$^m$18$^s$, DEC~(J2000)=--63\D54\arcmin29\arcsec. Also, we compared our \HII\ region detections with those found by \citet{1995ApJ...444..610R} by re-analysing his H$\alpha$ image. We found only 10 radio counterparts out of the 22 listed \HII\ regions in \citet{1995ApJ...444..610R} and show their properties in Table~\ref{hiiryder}. However, we note that \citet{1995ApJ...444..610R} selected these 22 \HII\ regions for spectroscopy to determine abundances. Apart from J190958-635405, another 6 radio \HII\ regions not selected/listed in \citet{1995ApJ...444..610R}  can be also detected in that H$\alpha$ image.

For 3 of these 17 \HII\ regions, we estimate radio spectral indices and find the values were slightly steeper than expected. However, they are still within the range --0.5 to +0.4. Interestingly, most of the bright IR sources within the \ngc\ are not detected in radio continuum. This is most likely because of our radio data sensitivity. However, as mentioned above, several radio features of the spiral arms are coincident with the IR counterparts.

%
\begin{table*}
\hspace*{-2cm}
\caption{17 radio continuum \ngc\ H\textsc{ii} regions found in this study. We list their positions, integrated flux densities and spectral indices where available.} \vspace{1mm}
\begin{center}
\begin{tabular}{ccccc}
\hline
  \multicolumn{1}{p{2.8cm}}{\centering H\textsc{ii} region\\ Name} &
  \multicolumn{1}{p{1.5cm}}{\centering RA~(J2000)  \\ (h m s)}              &
  \multicolumn{1}{p{1.5cm}}{\centering DEC~(J2000) \\ (\D \arcmin \arcsec)} &
  \multicolumn{1}{p{1.5cm}}{\centering S$_{2.1\,{\rm GHz}}$ \\ (mJy)} &
  \multicolumn{1}{p{1.9cm}}{\centering $\alpha \pm \Delta \alpha$} \\
\hline
J190919-635241 & 19:09:19.3 & --63:52:41.70 & 0.312 \\
J190920-635223 & 19:09:20.0 & --63:52:23.90 & 0.802 & --0.41$\pm$0.14 \\
J190922-635120 & 19:09:22.1 & --63:51:20.92 & 0.541 \\
J190931-634928 & 19:09:31.0 & --63:49:28.31 & 0.132 \\
\smallskip
J190932-635619 & 19:09:32.4 & --63:56:19.62 & 0.450 & --0.20$\pm$0.02 \\
J190943-635429 & 19:09:43.7 & --63:54:29.93 & 0.663 \\
J190950-635446 & 19:09:50.4 & --63:54:46.73 & 0.136 \\
J190952-635329 & 19:09:52.6 & --63:53:29.21 & 0.151 \\
J190955-635124 & 19:09:55.3 & --63:51:24.77 & 0.237 \\
\smallskip
J190956-634945 & 19:09:56.7 & --63:49:45.26 & 0.164 \\
J190958-635405 & 19:09:58.6 & --63:54:05.51 & 0.231 \\
J191004-635015 & 19:10:04.7 & --63:50:15.77 & 0.231 \\
J191008-635238 & 19:10:08.4 & --63:52:38.16 & 0.872 & --0.45$\pm$0.24\\
J191010-635218 & 19:10:10.6 & --63:52:18.35 & 0.400 \\
\smallskip
J191012-634931 & 19:10:12.4 & --63:49:31.82 & 0.080 \\
J191019-635532 & 19:10:19.5 & --63:55:32.78 & 0.428 \\
J191024-635430 & 19:10:24.0 & --63:54:30.83 & 0.680 \\
  \hline
\end{tabular}
\label{hii}
\end{center}
\end{table*}
%

%
\begin{table*}
\hspace*{-2cm}
\caption{22 H\textsc{ii} regions in \ngc\ found by \citet{1995ApJ...444..610R} in H$\alpha$. We estimate positions using the original H$\alpha$ image. Sources are matched to our radio catalog where applicable.} \vspace{1mm}
\begin{center}
\begin{tabular}{cccc}
\hline
  \multicolumn{1}{p{2.8cm}}{\centering \citet{1995ApJ...444..610R}\\ H\textsc{ii} Region Name} &
  \multicolumn{1}{p{2.0cm}}{\centering RA~(J2000)  \\ (h~ m~ s)}              &
  \multicolumn{1}{p{2.0cm}}{\centering DEC~(J2000) \\ (\D~ \arcmin~ \arcsec)} &
  \multicolumn{1}{p{2.7cm}}{\centering Radio Source Name} \\
\hline
s5a1 & 19:09:07.6 & --63:56:15 & \\
s2a4 & 19:09:20.0 & --63:52:24 & J190920-635223 \\
s2a2 & 19:09:22.1 & --63:51:21 & J190922-635120 \\
s5a2 & 19:09:27.1 & --63:56:19 & \\
\smallskip
s5a3 & 19:09:32.4 & --63:56:20 & J190932-635619 \\
s2a1 & 19:09:32.5 & --63:51:12 & \\
p2a4 & 19:09:42.2 & --63:54:59 & \\
p2a3 & 19:09:43.4 & --63:56:24 & \\
p2a5 & 19:09:50.4 & --63:54:47 & J190950-635446 \\
\smallskip
p2a7 & 19:09:51.9 & --63:52:40 & \\
p2a6 & 19:09:55.5 & --63:53:57 & \\
p1a1 & 19:09:57.3 & --63:50:37 & \\
s1a1 & 19:10:02.4 & --63:53:49 & \\
p1a2 & 19:10:04.7 & --63:50:16 & J191004-635015 \\
\smallskip
s1a2 & 19:10:04.9 & --63:53:24 & \\
s1a3 & 19:10:05.5 & --63:52:59 & \\
s1a4 & 19:10:08.4 & --63:52:38 & J191008-635238 \\
s1a5 & 19:10:10.6 & --63:52:18 & J191010-635218 \\
p1a4 & 19:10:12.4 & --63:49:32 & J191012-634931 \\
\smallskip
s4a1 & 19:10:19.5 & --63:55:33 & J191019-635532 \\
s4a2 & 19:10:24.0 & --63:54:31 & J191024-635430 \\
p1a5 & 19:10:26.5 & --63:49:32 & \\
  \hline
\end{tabular}
\label{hiiryder}
\end{center}
\end{table*}
%

%
%

\subsection{Infrared emission}
As described in \citet{2013AJ....145....6J}, the source (\ngc) characterisation consists of foreground star subtraction, background statistics, axis ratio and position angle determination at the 3$\sigma$ isophotal level, integrated flux and surface brightness measurements. The elliptical isophote shown in Figure~\ref{wiseimage2} corresponds to the 1$\sigma$ (RMS) isophote in the W1 (3.4~$\mu$m), which has a major axis diameter of 18.3\arcmin, equal to 50.4~kpc in physical size. This ellipse nicely matches D25 (the isophote of the 25.0 B magnitude per square arcsecond brightness level). The integrated flux down to this isophotal level is 1.98, 1.11, 3.59 and 4.20~Jy (with 1--2\% formal uncertainties), respectively for W1, W2, W3 and W4. The {\it WISE} flux density conversion values are 309.68, 170.66, 29.05 and 7.871~Jy respectively for W1, W2, W3 and W4\footnote{The native photometric calibration of WISE is the Vega system, where image measurements are computed in magnitudes and then converted to flux density (mJy) using the flux conversions from \citet{2011ApJ...735..112J}}. Extended emission below the single-pixel noise level is measured by fitting a double-Sersic to the axi-symmetric radial profile, which results in additional 3\% to 15\% emission, giving the ``total'' flux of the source: 1.99, 1.14, 4.23, 4.64~Jy, respectively.

Using the colour and integrated flux measurements, we derive the stellar mass and star formation properties. Following the procedure of \citet{2014ApJ...782...90C}, the W1 stellar mass-to-light ratio is estimated from the W1-W2=0.02~mag colour using integrated magnitudes of 5.488~mag and 5.466~mag for W1 and W2 respectively, giving a value of 0.59. Calculating the W1 inband luminosity from the total flux, assuming a distance of 9.5~Mpc, and applying the M/L ratio, leads to a stellar mass log(${\rm{M/M_{\odot}}}$) = 10.82$\pm$0.14. The star formation (SF) activity is derived from the W3 and W4 spectral luminosity ($\nu L_{\nu}$), where the stellar continuum has been subtracted based on the emission in the W1 band, and the total IR luminosity ($L_\mathrm{TIR}$) SFR relation from Cluver et al. 2017 (in prep). The resulting SFR is 4.7~${\rm{M_{\odot}/{yr}}}$ based on the W3 (12~$\mu$m) and 2.8~${\rm{M_{\odot}/{yr}}}$ based on the W4 (22~$\mu$m). The former includes contributions from PAH, nebular emission and warm dust, while the latter includes the warm dust continuum. This ``infrared'' SFR depends on the obscuration of dust to the young and massive (UV-emitting) stellar population, but note that it is very similar to the SFR based on emission from \HII\ regions, which is re-assuring for both infrared and H$\alpha$ estimates (stated previously in Section~\ref{sec:intro}).

Combining the two derived measures, the specific star formation rate (sSFR per Gyr) is then --10.18 to --10.37~yr$^{-1}$. The \ngc\ stellar mass and SFR are consistent with the predicted value based on the Galaxy SFR-M ``sequence'' that gauges the growth and evolution of galaxies \citep[see Fig.~13a in][]{2017ApJ...836..182J}. In terms of the specific SFR, which normalises the SFR across the mass spectrum (i.e. the sequence slope), the growth of \ngc\ is larger than that of late-type spirals (e.g. Milky Way and M\,31), but considerably less than starburst galaxies (e.g. M\,82), and not unlike that of the late-type barred M\,83, which has significant SF activity along the bar-ends \citep[see][]{2013AJ....145....6J}; and \citep[Fig.~13b in][]{2017ApJ...836..182J}. \ngc\ is still building its disk and bulge population, undergoing SF feedback processes, such as supernovae, that help to regulate its growth.

\subsection{Large-scale emission}

\begin{figure*}
 \begin{center} 
  \includegraphics[angle=-90, trim=0 0 0 0, width=0.7\linewidth]{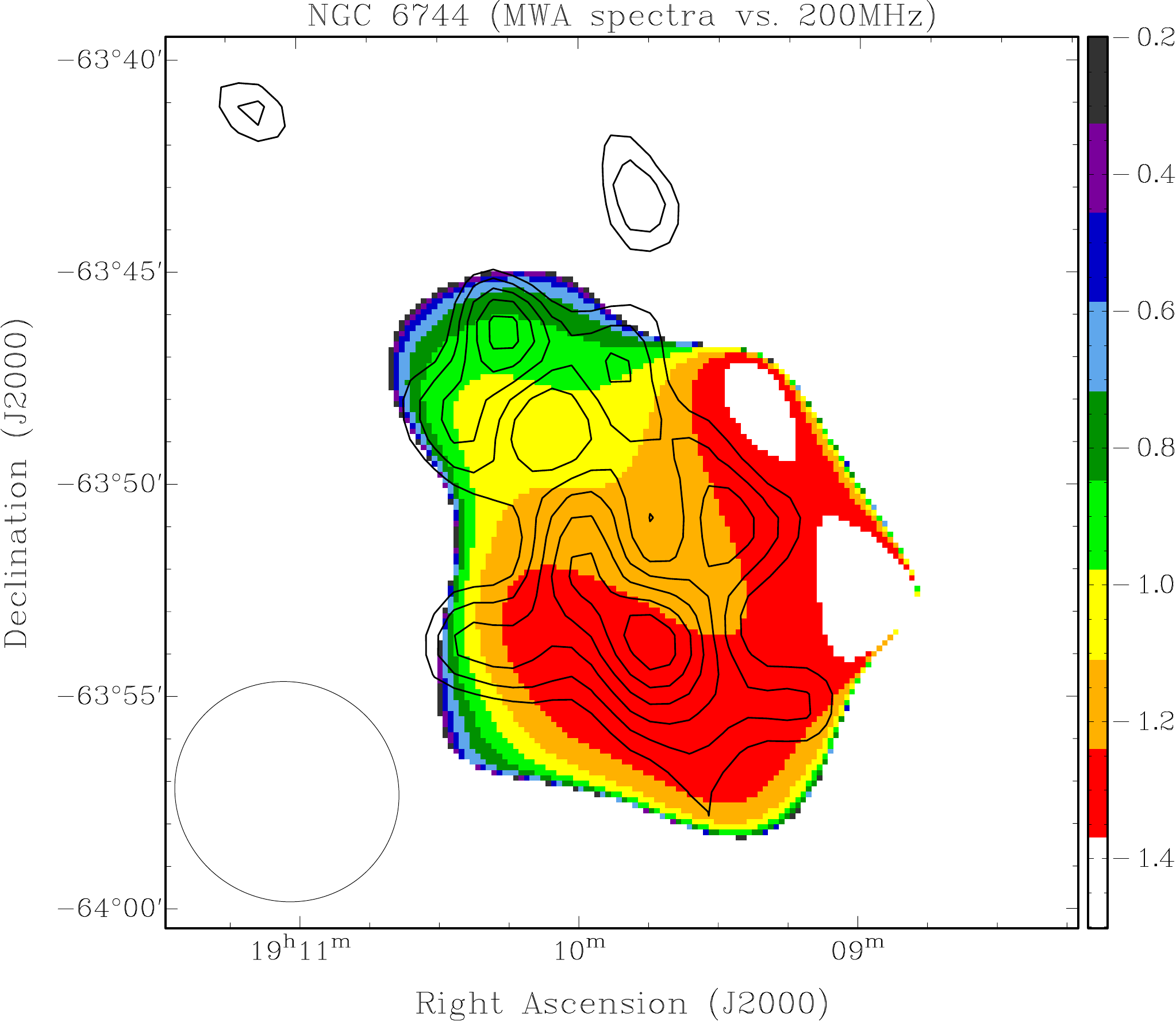}
 \caption{MWA spectral index map which is made of four MWA wide bands (88 -- 200~MHz) overlaid with 200~MHz contours. Contours are 4$\sigma$ to 10$\sigma$ with spacings of 1$\sigma$ (where 1$\sigma$=0.019~Jy~beam$^{-1}$). The synthesised beam size of the lowest MWA band (88~MHz) is shown in the lower left corner (5\arcmin$\times$5\arcmin). The color-bar indicates the spectral index value and is dimensionless.} 
 \label{wisemwaspec}
 \end{center}
\end{figure*}

A low-frequency MWA image of \ngc\ \citep[from the GLEAM survey,][]{2015PASA...32...25W} is compared to the 3-band {\it WISE} image in Figure~\ref{wisemwa}; a low-frequency spectral index image, based on all four wide MWA bands centred on 88~MHz, 118~MHz, 154~MHz and 200~MHz, is shown in Figure~\ref{wisemwaspec}. The synchrotron-dominated 200~MHz emission is elongated NE--SW, and has a greater extent than the N--S elongated IR emission. The northeast end is dominated by a strong (most likely background) source as well as a similar extension to the southeast. It therefore appears that the large-scale synchrotron emission from \ngc\ is somewhat more extended than, and roughly aligned with, the IR emission from its disk.

Figure~\ref{wisemwaspec} shows a spectral index gradient along the elongation axis of the synchrotron emission. The spectral index is steeper in the SW region and shallower in the NE ($\alpha \sim -0.7$). While the errors associated with the spectral index are large ($\sim 0.4$) and the index is calculated over a frequency range limited to lower frequencies, we still find this ``bipolar'' spectral index very unusual and further study at other radio frequencies are needed.

%
%

\section{CONCLUSION}
 \label{sec:conclusion}

We present a catalogue of the 387 discrete X-ray and radio sources in \ngc\ which are shown here for the first time. Our multi-frequency analysis of the available observations have led to the following results:
\begin{itemize}
 \item We find a central source which we associate with a supermassive black hole (AGN). It is detected in both X-ray and radio, though very faint, with possible jet flows spanning in the SE to NW direction. We estimate the bolometric radio luminosity as $\sim1.3\times10^{36}$~ergs~s$^{-1}$ which is an order of magnitude stronger compared to the Milky Way's central black hole ($\sim$10$^{35}$~ergs~s$^{-1}$). Similarly, the X-ray luminosity of the central source is a few orders of magnitude larger in \ngc\ than in our Galaxy as expected based on previous studies.
 \item We find a total of 117 X-ray sources including one foreground star that are identified here for the first time. Also, we find 280 radio sources from which 254 sources are categorised as background objects. Of these, we identified nine sources in common between X-ray and radio as well as only two common to all three wavebands. We find the minimum detectable flux of the X-ray sources as 8.23$\times10^{-16}$~ergs~s$^{-1}$~cm$^{-2}$.
 \item We confirm 2 sources as likely SNRs, and identify a further 5 SNR candidates. Given the unusually large fluxes recorded by \chandra\ at a distance of 9.5~Mpc, we suggest that the SNRs are particularly bright. However, the new SNRs are not as bright as the well-known Galactic precedent, SNR Cassiopeia~A \citep{1996ApJ...466..309K,2007RMxAC..30...12R}.
 \item We find 17 \HII\ regions, with two clearly in the spiral arms of the \ngc. {\it WISE} data show relatively cool material in the outer spiral arms of the galaxy. This 22-$\mu$m dominated emission is likely to arise from the warm dust in the vicinity of \HII\ regions.
 \item From {\it WISE} data, we estimate the SFR to be 2.8--4.7~$\rm{M_{\odot}/{yr}}$. This value along with the stellar mass of \ngc\ are consistent with the SFR and population evolutionary growth predicted for an intermediate disk galaxy \citep[e.g., see][]{2013AJ....145....6J}. For all the apparent similarities between \ngc\ and our Milky Way, the global SFR in \ngc\ is 2-3 times greater than that of the Milky Way. 
\end{itemize}

\begin{acknowledgements}
The Australia Telescope Compact Array is part of the Australia Telescope National Facility which is funded by the Commonwealth of Australia for operation as a National Facility managed by CSIRO. 
This research has made use of data obtained from the \chandra\ Data Archive and the \chandra\ Source Catalog and software provided by the \chandra\ X-ray Center (CXC) in the application packages CIAO, ChIPS, and Sherpa.
ADK acknowleges financial support from the ARC Centre of Excellence for All-Sky Astrophysics (CAASTRO) through project number CE110001020. 
We thank an anonymous referee for their thorough review and highly appreciate their comments and suggestions, which significantly contributed to improving the quality of our paper. 
We thank Stuart Ryder for providing the H$\alpha$ image from his earlier study \citep{1995ApJ...444..610R}.
MZP acknowledges support from the Ministry of Education, Science and Technological Development of the Republic of Serbia through project No. 176005.
\end{acknowledgements}

\nocite*{}
\bibliographystyle{pasa-mnras}
\bibliography{References}

\end{document}